\documentclass[12pt]{article}
\usepackage{graphicx}
\begin{document}
\baselineskip=16pt
\title{ \addvspace{-10mm}
 \hfill {\normalsize CFNUL/98-08} 
\\  
\addvspace{15mm}
Neutrino Oscillations: a source of Goldstone fields\\ and consequences for
Supernovae\footnote{Talk given at the International Symposium on Lepton and
Baryon Number Violation, Trento, Italy, 1998}\\ }
\author{Lu\'{\i }s Bento \\
\em  Centro de F\'{\i }sica Nuclear da Universidade de Lisboa,\\
\em  Av. Prof. Gama Pinto 2, 1699 Lisboa - {\sl codex}, Portugal\\
{\rm e-mail: lbento@fc.ul.pt}}

\date{}

\maketitle

\begin{abstract}
It is shown that true Nambu-Goldstone (NG) bosons develop coherent fields
whenever the associated charges of the matter particles are not conserved in
a macroscopic scale. The sources of the NG fields are the time rates of
quantum number violation. If the lepton numbers are spontaneously broken at
a scale below 1 TeV, the neutrino oscillation processes generate classic NG
fields that are strong enough in Supernovae to modify the neutrino flavor
dynamics.
The oscillation patterns may change in the periods of largest $\nu $ fluxes.
Two examples are given: 1. the back reaction of a NG field improves the
adiabaticity of the $\nu _{e}$ resonant conversion; 2. ${\bar{\nu}_{e}}%
\leftrightarrow {\bar{\nu}_{\mu }}$ oscillations may occur even if $\nu _{e}$
is the lightest of the neutrinos.
\end{abstract}





\newpage

\section{Basic Features}

The Goldstone\ theorem asserts that \strut in a theory with spontaneous
breaking of a global symmetry group some massless scalar bosons should exist
- the so-called Nambu-Goldstone bosons (NG) - one per global symmetry broken
by the vacuum. Good candidates for spontaneously broken quantum numbers are
the partial lepton numbers $L_{e},L_{\mu },L_{\tau }$ and total lepton
number $L$. Indeed, they are conserved by the Standard Model (SM)
interactions, but several experiments going now from solar to atmospheric
neutrinos and laboratory oscillation experiments \cite{cald98} indicate that
the lepton flavors are not conserved.

The NG bosons not only have zero mass but also no scalar potential terms
such as $\phi ^{4}$. Futhermore, they also interact with the fermion
particles however, they do not mediate long range forces. The reason lies in
the very global symmetry that, broken by the vacuum still operates at the
Lagrangian level. A NG boson transforms as $\phi \rightarrow \phi +\alpha
\,, $ $\alpha ={\rm constant}$, and therefore the symmetry requires that
these fields have nothing but derivative couplings (this is related to the
soft pion low energy theorems \cite{chen89,wein96}) {\em i.e.}, all of the
NG boson interactions can be expressed with an effective Lagrangian of the
form 
\begin{equation}
{\cal L}={\frac{1}{2}}\,\partial _{\mu }\phi \,\partial ^{\mu }\phi +{\cal L}%
_{{\rm int}}(\partial _{\mu }\phi )\,.  \label{Luni}
\end{equation}
The $\phi $ equation of motion is nothing but the conservation law of the
total current associated with the $\phi \rightarrow \phi +\alpha \,$
symmetry, 
\begin{equation}
\partial _{\mu }\partial ^{\mu }\,\phi =-\partial _{\mu }J_{\Lambda }^{\mu
}/V_{\Lambda }\,,  \label{ddfi}
\end{equation}
where $V_{\Lambda }$ is essentially the scale of symmetry breaking and the
current $J_{\Lambda }^{\mu }$ is determined in leading order by the quantum
numbers of the particles. For instance, if $\Lambda $ is the electron lepton
number $L_{e}$, 
\begin{equation}
J_{e}^{\mu }={\bar{e}\,}\gamma ^{\mu }\,e+{\bar{\nu}_{e}}\gamma ^{\mu }\nu
_{e}+\cdots \,,  \label{je}
\end{equation}
where the dots stand for model dependent radiative corrections and scalar
boson contributions.

This general result can be obtained using a suitable rephasing of the fields
present in the theory \cite{gelm83,bent97}. Let $\Lambda _{a}$ and $\Lambda
_{i}$ be the quantum numbers of the fermion fields $\chi ^{a}$ and scalars $%
\sigma _{i}$ under an abelian global symmetry${\rm {\ U(1)_{\Lambda }}}$
which is spontaneously broken by vacuum expectation values $\left\langle
\sigma _{i}\right\rangle $. A suitable change of variables namely, 
\begin{eqnarray}
\chi ^{a} &=&e^{-i\,\Lambda _{a}\phi /V_{\Lambda }}\psi ^{a}\ ,\smallskip
\label{tpsi} \\
\sigma _{i} &=&e^{-i\,\Lambda _{i}\phi /V_{\Lambda }}\,\left( \left\langle
\sigma _{i}\right\rangle +\rho _{i}\right) \ ,  \label{tsigma}
\end{eqnarray}
makes the Lagrangian to be expressed in terms of {\em physical weak}
eigenstates, fermions $\psi ^{a}=e,\,\nu _{e},\,...$, massive bosons $\rho
_{i}$ and gauge bosons (assumed to be singlets of the global symmetry) all
of them invariant under U(1)$_{\Lambda }$. The only non-invariant field is
the NG boson $\phi $. It transforms as\ $\phi \rightarrow \phi +\alpha $
hence, the Lagrangian, symmetric under U(1)$_{\Lambda }$, can only depend on 
$\phi $ through the derivatives $\partial _{\mu }\phi $. Its leading order
couplings are derived from the kinetic Lagrangians of $\chi ^{a}$ and $%
\sigma _{i}$ by application of Eqs.\ (\ref{tpsi}-\ref{tsigma}). One obtains in
particular, $V_{\Lambda }^{2}=2\sum |\Lambda _{i}^{{}}\left\langle \sigma
_{i}\right\rangle |^{2}$.

\strut The second member of the NG equation of motion, Eq.\ (\ref{ddfi}), is
clearly not an ordinary scalar density. If one calculates the divergency of
a vector or axial-vector current using the Dirac equation, one ends up with
either scalar but flavor violating terms,

\[
\partial _{\mu }\bar{\psi}_{i}\gamma ^{\mu }\psi _{j}=i(m_{i}-m_{j})\bar{\psi%
}_{i}\psi _{j}\ , 
\]
or flavor conserving but pseudo-scalar,

\[
\partial _{\mu }\bar{\psi}_{i}\gamma ^{\mu }\gamma _{5}\psi
_{j}=i(m_{i}+m_{j})\bar{\psi}_{i}\gamma _{5}\psi _{j}\ . 
\]
This kind of source terms do not generate long range '$1/r$' fields, but
rather spin dependent '$1/r^{3}$' potentials \cite{chik81}.

In the intent of escaping the derivative couplings law, models were made
with explicit symmetry breaking terms. These might be anomaly terms \cite
{wilc82}, as in the axion case \cite{pecc77}, or soft symmetry breaking
terms \cite{hill88}. In this way scalar couplings with the NG field are
produced but at the cost of giving a mass to the then, so-called pseudo
Nambu-Goldstone boson. The mass ultimately means a finite range for the NG
bosons. I will show next that under certain conditions a true
Nambu-Goldstone\ boson may develop a macroscopic field \cite{bent97,bent98}.

\section{Long range Nambu-Goldstone fields}

First notice that the source of a NG field is different from zero if
the matter current $J_{\Lambda }^{\mu }$ is not conserved. The volume
integral

\[
\int \!{d^{3}x\,\ }\partial _{\mu }J_{\Lambda }^{\mu }=\frac{d\Lambda }{dt}^{%
\hspace{-2.83pt}{\rm created}} 
\]
represents the rate of $\Lambda $-number {\em created} per unity of time (in
all particles except $\phi $) and $\partial _{\mu }J_{\Lambda }^{\mu }$ is
its volume density. The equation of motion (\ref{ddfi}) gives 
\begin{equation}
\phi (t,\vec{r})={\frac{{-1}}{{V_{\Lambda }}}}\int \!{d^{3}x\,{\frac{\dot{%
\rho}_{\Lambda }^{{\rm cr}}(t-{\left| {\vec{r}-\vec{x}}\right| },\vec{x})}{{%
4\pi \left| {\vec{r}-\vec{x}}\right| }}}}\;.  \label{field}
\end{equation}
It is a long-range Coulomb type of field but suppressed by the product of
the scale ${V_{\Lambda }}$ times the interval of time per $\Lambda $%
-violating process. To have a significant macroscopic field one needs a
large rate of reactions with a net $\Lambda $ number variation. That calls
for astrophysical sources. There is evidence that solar neutrinos change
flavor as they come out of the Sun. If $\nu _{e}$ oscillates into ${\nu
_{\mu }}$, for instance, then ${L_{e}}$ and ${L_{\mu }}$ are not conserved
in a large scale in stars and possibly in Supernovae as well. If any of them
is associated with a spontaneously broken global symmetry then, the
respective NG boson acquires classic field configurations.

For definiteness suppose that ${\,}L_{e}$ is spontaneously broken and that a
fraction of the $\nu _{e}$s emitted from a star are resonantly converted
into $\nu _{\mu }$ through the Mikheyev-Smirnov-Wolfenstein (MSW) mechanism 
\cite{wolf78,mikh86} (also $\nu _{\mu }$ into $\nu _{e}$ in a Supernova) in
a certain shell of radius $\sim r_{{\rm c}}$ inside the star. Assuming for
simplicity stationary neutrino fluxes and spherical symmetry one obtains the
solution for the NG boson $\phi _{e}$:
\begin{eqnarray}
 \phi _{e} &=& {\rm const.}\ ,\quad r<r_{{\rm c}}\, ,
\nonumber \\ 
&  &  \nonumber \\
 \phi _{e} &=&\frac{1}{V_{\Lambda }}\frac{1}{4\pi \,r}\left[ 
\frac{dN}{dt}({\nu _{e}\to \nu _{\mu })-}\frac{dN}{dt}({\nu _{\mu }\to \nu
_{e})}\right] \ ,\quad r>r_{{\rm c}}
\end{eqnarray}
in terms of the total numbers of converted
neutrinos, $N(\nu _{\mu }\to \nu _{e})$ and $N(\nu _{e}\to \nu _{\mu })$,
per unity of time.

\strut What are the possible consequences of such a NG field? Applying Eqs.\ (%
\ref{tpsi}) to the kinetic terms

\[
{\cal L}_{{\rm 0}}=\overline{{e}}\,i\,\partial \hspace{-0.22cm}/\,e+%
\overline{\nu }_{e}\,i\,\partial \hspace{-0.22cm}/\,\nu _{e}\ , 
\]
one obtains the interaction

\begin{equation}
{\cal L}_{{\rm int}}=\frac{1}{V_{\Lambda }}\partial _{\mu }\phi _{e}\,\left( 
\overline{{e}}\,\gamma ^{\mu }\,e{+}\overline{\nu }_{e}\,\gamma ^{\mu }\,\nu
_{e}\right) \,.  \label{lint}
\end{equation}
The fermions only care about the gradient of the NG field. In the situation
described above, denoting by $\vec{j}({\nu _{e}\to \nu _{\mu })}$ the flux
of $e$-neutrinos converted to ${\nu _{\mu }}$ and $\vec{j}({\nu _{\mu }\to
\nu _{e})}$ the reciprocal, 
\begin{equation}
\vec{A}_{e}=-\frac{\vec{\nabla}\phi _{e}}{V_{\Lambda }}=\frac{1}{V_{\Lambda
}^{2}}\left[ \vec{j}({\nu _{e}\to \nu _{\mu })}-\vec{j}({\nu _{\mu }\to \nu
_{e})}\right] \ .  \label{Ae}
\end{equation}
The quantity $\vec{A}_{e}$ is suppressed relatively to the field $\phi _{e}$
by the scale $V_{\Lambda }$ and radius $r$ and that is what makes it
extremely small. However, as follows from Eq.\ (\ref{lint}), it still
contributes to the potential energy of an electron-neutrino with velocity $%
\vec{v}_{\nu }$ as 
\begin{equation}
V_{\nu _{e}}=\vec{A}_{e}\cdot \vec{v}_{\nu }\;.  \label{Vnue}
\end{equation}
These contributions associated with breaking of partial lepton numbers are
flavor dependent by nature and therefore may have a role in neutrino
oscillations.

One estimates the order of magnitude of theses NG potentials as the neutrino
flux divided by $V_{\Lambda }^{2}$. In the case of the Sun, evaluating the
total flux at the Sun radius $R_{\odot }=7\times 10^{10}\,{\rm cm}$, 
\[
V_{{\rm NG}}\sim \frac{j_{\nu }}{V_{\Lambda }^{2}}=\frac{2.8\times 10^{-2}}{%
R_{\odot }}\frac{1\,{\rm keV}^{2}}{V_{\Lambda }^{2}}\ . 
\]
It produces a too small variation of the neutrino phase, $\sim V_{{\rm NG}%
}\,R_{\odot }$, thus it does not seem significant for solar neutrinos. In a
Supernova however the fluxes are many orders of magnitude larger. One has 
\begin{equation}
 \frac{j_{\nu }}{V_{\Lambda }^{2}}=1.48\ \frac{G_{\Lambda }}{%
G_{{\rm F}}}\frac{L_{\nu }}{10^{52}\,{\rm ergs/s}}\,\frac{10\,{\rm MeV}}{%
\left\langle E_{\nu }\right\rangle }\left( \frac{r}{10^{10}\,{\rm cm}^{{}}}%
\right) ^{-2}\times {}10^{-12}\,\,{\rm eV\ ,}  \label{VNG}
\end{equation}
where $G_{\rm{F}}=11.66\,{\rm TeV}^{-2}$ is the Fermi constant, $%
G_{e}=1/V_{\Lambda }^{2}$, $L_{\nu }$ is the neutrino energy luminosity, and 
$\left\langle E_{\nu }\right\rangle $ its average energy. $L_{\nu
}/\left\langle E_{\nu }\right\rangle $ gives the neutrino emission rate.
This has to be compared with the Standard Model electro-weak potential $%
V_{W} $ \cite{wolf78}. In the regions of a Supernova star with densities
typical of the Sun the mass density goes as $1/r^{3}$, the constant $\tilde{M%
}=\rho \,r^{3}$ lying between $10^{31}{\rm g}$ and $15\times 10^{31}{\rm g}$
depending on the star \cite{wils86}. Then, 
\begin{equation}
V_{W}=\sqrt{2}{\rm \,}G_{{\rm F}}\,n_{e}=0.76\ Y_{e}{\frac{{\tilde{M}}}{%
10^{31}{\rm g}}\,}\left( \frac{r}{10^{10}\,{\rm cm}^{{}}}\right) ^{-3}\times
10^{-12}\,\,{\rm eV}\;,  \label{VW}
\end{equation}
$Y_{e}\approx 1/2$ being the electron abundance and $n_{e}$ the number
density. Clearly, in the period of time where the neutrino luminosities are
high enough say, $10^{52}\,{\rm ergs/s}$, the scale of lepton symmetry
breaking can be as high as the Fermi scale and yet the NG potentials compete
with $V_{W}$ at the places where the resonance occurs for values of $\Delta
m^{2}$ interesting for solar neutrino solutions ($\Delta m^{2}/E\sim
10^{-12}\,{\rm eV}^{2}$). Futhermore, the NG potentials decay as $1/r^{2}$
and necessarily overcome the local interactions at large enough radius.

At this point we can already draw some conclusions. First, if the lepton
numbers $L_{e},L_{\mu },...$ are spontaneously broken, the neutrino
oscillations in a Supernova originate the appearance of classic
Nambu-Goldstone fields. Second, if the scale of symmetry breaking lies below
1 TeV they in turn yield potential energies that are significant for the
very neutrino oscillation processes. What are the possible consequences? The
NG coupling are flavor dependent by nature hence, may alter the pattern of
neutrino oscillations, with a fundamental difference: the NG potentials are
proportional to the neutrino fluxes and that induces a time dependence of
the effects in a Supernova context.

Next I show two examples. One \cite{bent98} is the improvement of
adiabaticity of otherwise $\nu _{e}\leftrightarrow \nu _{X}$ non-adiabatic
transitions. The other \cite{bent97} is the violation of the SM prediction
that the anti-neutrinos do not resonantly oscillate if the neutrino masses
obey the hierarchy of the charged lepton masses.

\section{Improvement of adiabaticity: Majoron back reaction}

Suppose that in the Sun the electron-neutrino oscillates into the
muon-neutrino with the parameters of the non-adiabatic, small mixing angle
solution \cite{haxt86,park86,rose86}. Then, the $\nu _{e}\leftrightarrow \nu
_{\mu }$ transitions are also non-adiabatic in a Supernova and only a
fraction of each neutrino species is converted into the other. Furthermore,
since the level crossing probability is an increasing function of the
energy, the hotter $\nu _{\mu }$s\ have larger survival probabilities than
the cooler $\nu _{e}$s. Let us assume that the partial lepton number $L_{e}$
is conserved at the Lagrangian level but the respective global symmetry, U(1)%
$_{e}$, is spontaneously broken. Then, a NG boson (Majoron), $\phi _{e}$,
exists with zero mass. Although in principle, the neutrino mass matrix
violates all three lepton numbers, I will ignore for simplicity other
possibly existing Majorons. Notice that other NG fields are naturally less
intense if the respective scales of symmetry breaking are slightly higher.

\strut The equations governing two flavor oscillation \cite{kuo89} with
mixing angle $\theta $ and $\Delta m^{2}$ difference in the vacuum,\ are
given by, up to an universal phase absorbed in the wave functions, 
\begin{equation}
i\frac{\partial }{\partial \,r}\left( 
\begin{array}{c}
{\nu _{e}} \\ 
\\ 
{\nu _{\mu }}
\end{array}
\right) =\frac{1}{2E}\left( 
\begin{array}{cc}
2E\,V_{e} & \quad {\rm \,}\frac{1}{2}\Delta m^{2}\sin 2\theta \\ 
&  \\ 
{\rm \,}\frac{1}{2}\Delta m^{2}\sin 2\theta & \quad {\rm \,\,}\Delta
m^{2}\cos 2\theta
\end{array}
\right) \left( 
\begin{array}{c}
{\nu _{e}} \\ 
\\ 
{\nu _{\mu }}
\end{array}
\right) \;.  \label{osc1}
\end{equation}
$V_{e}$ is the difference between the ${\nu _{e}}$ and ${\nu _{\mu }}$
potentials, which is the following in the presence of a $\phi _{e}$ field
(see Eqs.\ (\ref{Vnue},\ref{VW}): 
\begin{equation}
V_{e}=V_{{\nu _{e}}}-V_{{\nu _{\mu }}}=V_{W}+\vec{A}_{e}\cdot \vec{v}_{\nu
}\ .  \label{DV1}
\end{equation}
Assuming spherical symmetry, $\vec{A}_{e}$ only has radial component and one
obtains from Eq.\ (\ref{Ae}) 
\begin{equation}
V_{e}=\sqrt{2}\,G_{{\rm F}}\,n_{e}+\frac{G_{e}}{4\pi \,r^{2}}\left[ \dot{N}({%
\nu _{e}\to \nu _{\mu })-}\dot{N}({\nu _{\mu }\to \nu _{e})}\right] \ .
\label{DV2}
\end{equation}
\ \ $V_{e}$ would be the same if one considers alternatively a NG boson
associated with the breaking of $L_{\mu }$ or $L_{e}-L_{\mu }$. The reason
is, $\nu _{e}\leftrightarrow \nu _{\mu }$ oscillations only violate $%
L_{e}-L_{\mu }$ not the total number $L_{e}+L_{\mu }$.

In order to calculate the radial component $A_{e}(r)$\ one needs to know the
fluxes of $\nu _{\mu }$ and $\nu _{e}$\ as functions of the radius and their
energy spectra. Let $f_{\nu _{e}}(E)$ and $f_{\nu _{\mu }}(E)$ be the number
of neutrinos emitted per unity of time and energy, ${\rm d}\dot{N}_{\nu
_{e}}/{\rm d}E$ and ${\rm d}\dot{N}_{\nu _{\mu }}/{\rm d}E$ respectively. If
the transition is non-adiabatic, the number of converted neutrinos with a
given energy $E$ is only a fraction $1-P_{c}$\ of the radiated particles, $%
P_{c}$ is the level crossing probability, and so 
\begin{eqnarray}
d\dot{N}({\nu _{e}}{\to \nu _{\mu })} &= (1-P_{c})\,f_{\nu _{e}}(E)\,dE\
,\medskip  \label{dNe} \\
d\dot{N}({\nu _{\mu }}{\to \nu _{e})} &= (1-P_{c})\,f_{\nu _{\mu
}}(E)\,dE\ .  \label{dNmu}
\end{eqnarray}
The problem of obtaining these rates as functions of the radius is
simplified by neglecting the oscillation length of the resonantly converted
neutrinos: in that limit a neutrino with energy $E$ is converted to the
other flavor at the position where the resonance condition is verified: 

\begin{equation}
E(r)=\frac{\Delta m^{2}\cos 2\theta }{2\,V_{e}(r)}\ .  \label{ER}
\end{equation}
In addition, the level crossing probability $P_{c}$\ can be calculated using
the Landau-Zener approximation \cite{haxt86,kuo89},

\begin{equation}
P_{c}(E)=\exp \left\{ -\frac{\pi }{4}\left| \frac{{\rm \,}\Delta m^{2}}{{\rm %
d}E/{\rm d}r}\frac{\sin ^{2}2\theta }{\cos 2\theta }\right| \right\} _{\!\!%
\rm{res.}}\ .  \label{Pc}
\end{equation}
These approximations change the numbers but not, I believe, the lesson taken
by comparing the results with and without a Majoron field. Eqs.\ (\ref{VW},%
\ref{DV2} - \ref{Pc}) establish the differential equations for $d\dot{N}({%
\nu _{e}\to \nu _{\mu })}$ and $d\dot{N}({\nu _{\mu }\to \nu _{e})}$. They
are non-linear because the derivative of $E(r)$\ is not independent from the
Majoron potential $A_{e}(r)$.

\begin{figure}[t]
\centering
\includegraphics*[width=80mm]{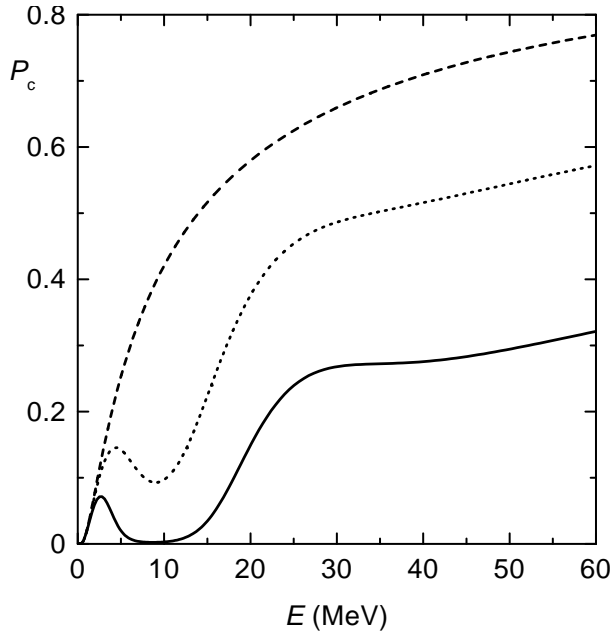}
\caption{Level crossing probability as a function of the $\nu $
energy. Dashed curve: SM with $\tilde{M}=4\times
10^{31}\,{\rm g}$\ and $Y_{e}=1/2$. Dotted and bold curves: a
Nambu-Goldstone field exists with $G_{e}=G_{{\rm F}}$ and $%
G_{e}=4\,G_{{\rm F}}$ respectively. The luminosities are $10^{52}\,{\rm %
ergs/s}$ for $\nu _{e}$ and $7\times 10^{51}\,{\rm ergs/s}$ for $\nu _{\mu }$%
.}
\end{figure}

Let us examine the $\nu _{e}\leftrightarrow \nu _{\mu }$\
oscillations with the mixing parameters of the non-adiabatic solar neutrino
solution (for a recent update see \cite{hata97}), choosing in particular the
values $\Delta m^{2}=7\times 10^{-6}\,{\rm eV}^{2}$, $\Delta m^{2}\sin
^{2}2\theta =$\ $4\times 10^{-8}\,{\rm eV}^{2}$. In a Supernova the
resonance is non-adiabatic as well and, as Eq.\ (\ref{Pc}) indicates, the
survival probability $1-P_{c}$ increases with the $\nu $ energy. A detailed
study in the framework of the SM was done in \cite{mina88}. In Fig.\ 1, $%
P_{c} $ in the Landau-Zener approximation is plotted against the energy. The
dashed curve holds for the SM potential with a constant $\tilde{M}=4\times
10^{31}{\rm g}$. It is manifest the aggravation of the non-adiabaticity with
the energy.

To study the Majoron case one has to specify the energy spectra and
luminosities. I used Fermi-Dirac distributions with the following values of
temperature and chemical potential \cite{burr90}: for $\nu _{e}$, $T=2.4{\rm %
\ MeV}$ and$\ \mu =3.2\,T$; for $\nu _{\mu }$, $T=5.1{\rm \ MeV},\ \mu
=4.1\,T$. This gives average energies of $10$ and $23{\rm \ MeV}$
respectively. The luminosity intensities are in turn $10^{52}\,{\rm ergs/s}$
for $\nu _{e}$ and $7\times 10^{51}\,{\rm ergs/s}$ for $\nu _{\mu }$ which
amount to particle emission rates of $10^{51}$ and $3\times 10^{50}\,{\rm %
ergs/s/MeV}$ respectively. Because the $e$-neutrinos are more numerous, the $%
\nu _{e}\leftrightarrow \nu _{\mu }$\ oscillations produce a net destruction
of $L_{e}$-number and a positive potential $A_{e}$. The dynamics is the
following: the less energetic $\nu _{e}$\ oscillate to $\nu _{\mu }$\ at
smaller radius than $\nu _{\mu }\rightarrow \nu _{e}$; the $\nu
_{e}\rightarrow \nu _{\mu }$ conversion produces a positive $A_{e}(r)$\
which attenuates the fall of the total potential $V_{e}=V_{W}+A_{e}$ with $r$%
; Consequently, the adiabaticity improves at larger radius and the most
energetic neutrinos change flavor with higher probabilities.

\begin{figure}[t]
\centering
\includegraphics*[width=80mm]{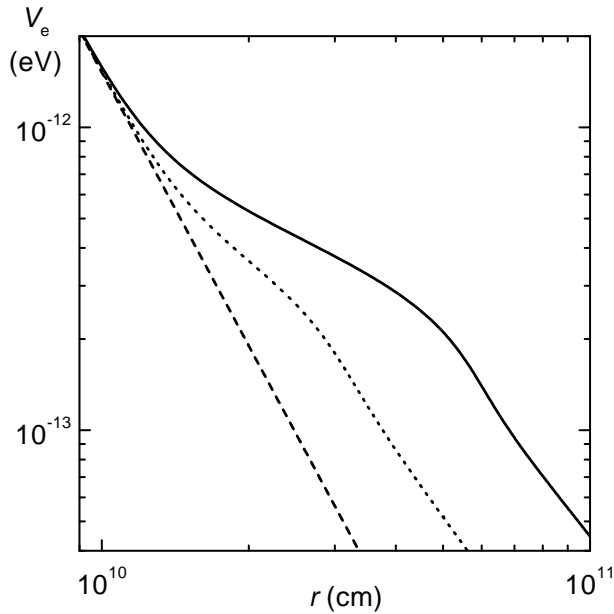}
\caption{Total potential $V_{e}=V_{\nu _{e}}-V_{\nu _{\mu }}$\ as a function
of the radius. The dashed curve is the SM potential, the dotted and
bold curves are the Majoron cases with the same parameters as in Fig. 1.}
\end{figure}

Fig.\ 2 \cite{bent98} shows the total potential $V_{e}$\ as a function $r$.
The dashed curve stands for the SM potential, $V_{W}$, with $\tilde{M}%
=4\times 10^{31}\,{\rm g}$\ and $Y_{e}=1/2$. The dotted and bold curves
differ in the scale of $L_{e}$ symmetry breaking, $G_{e}=G_{{\rm F}}$ and $%
G_{e}=4\,G_{{\rm F}}$, respectively. The potential falls more slowly as
the field $\phi _{e}$ grows which reflects on the level crossing probability
plotted in Fig.\ 1 \cite{bent98} for the same cases (dotted and bold
curves). The effect is clear: the stronger the Majoron field, the more
efficient is the flavor conversion. It is worth to mention that if the $\mu $%
-neutrinos were more numerous than the $e$-neutrinos the effect would be the
opposite because the Majoron potential would be negative ($\dot{L}_{e}>0$).
That shows up in the rise of $P_{c}$\ at the $\nu _{\mu }$\ energy band
around $20{\rm \ MeV}$.

\begin{figure}[t]
\centering
\includegraphics*[width=3.375in]{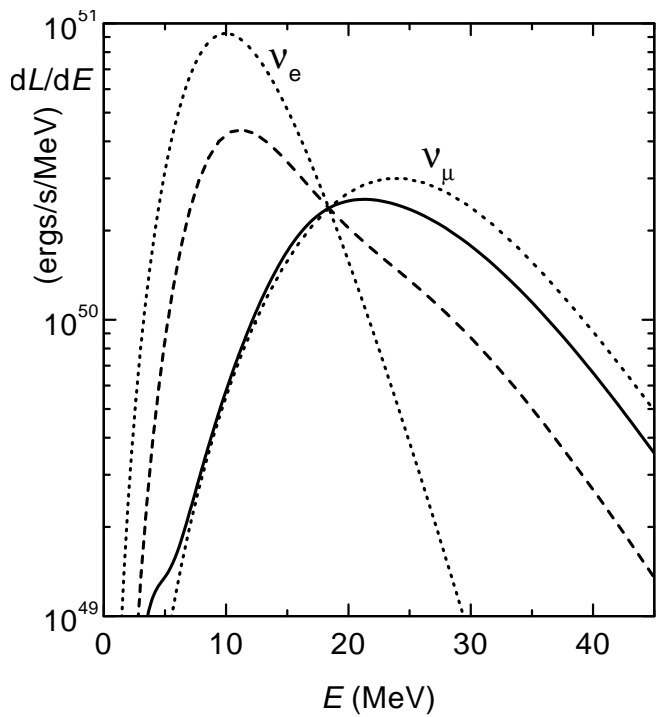}
\caption{ The dotted curves are the assumed luminosity distributions for $%
\nu _{e}$ and\ $\nu _{\mu }$ as emitted from the neutrinospheres. The dashed
curve represents the $\nu _{e}$ luminosity after electroweak neutrino
oscillations whereas in the bold curve a $\phi _{e}$ field exists with $%
G_{e}=4\,G_{{\rm F}}$.}
\end{figure}

Fig.\ 3 \cite{bent98} shows the implications for the outgoing $\nu _{e}$\
energy spectrum. The dotted curves are the assumed luminosity distributions
for the emitted $\nu _{e}$s and\ $\nu _{\mu }$s. The dashed curve represents
the luminosity of the $e$-neutrinos that come out of the star after standard
MSW oscillations and the bold curve is the same but with a Majoron field ($%
G_{e}=4\,G_{{\rm F}}$). The improvement of adiabaticity makes more $\nu
_{\mu }$s convert into $\nu _{e}$\ and less $\nu _{e}$\ to survive, and
because the $\mu $-neutrinos are more energetic, the outgoing $\nu _{e}$\
spectrum is harder than if there was no Majoron field. The average energy of
the outgoing $e$-neutrinos\ is $13{\rm \ MeV}$ for $G_{e}=0$ but reaches $17%
{\rm \ MeV}$ if $G_{e}=G_{{\rm F}}$ and $21{\rm \ MeV}$\ if $G_{e}=4\,G_{%
{\rm F}}$. These effects on the spectra can in principle be traced in those
detectors such as Super-Kamiokande and SNO, capable of detecting Supernova
electron-neutrinos \cite{burr92}.

To summarize, if the explanation of the solar neutrino deficit is the
MSW non-adiabatic oscillation $\nu _{e}\rightarrow \nu _{\mu }\ $($\nu
_{e}\rightarrow $ $\nu _{\tau }$ or any other superposition of $\nu _{\mu }$
and $\nu _{\tau }$) then, the Standard Model predicts that in a Supernova
the $\nu _{e}\leftrightarrow \nu _{\mu }$\ transitions are also
non-adiabatic. It means that, to a large extent, the $e$-neutrinos preserve
their lower energy spectrum, unless $\nu _{e}$\ also mixes to another flavor
with a too high or too low $\Delta m^{2}$\ to show up in solar neutrinos. If
however, $L_{e}$\ is a spontaneously broken quantum number, the associated
Nambu-Goldstone boson, $\phi _{e}$, will acquire a classic field
configuration which may be strong enough to produce a back reaction that
improves the adiabaticity of the $\nu _{e}\leftrightarrow \nu _{\mu }$\
transitions. The outcome is a $\nu _{e}$\ energy spectrum harder than
expected. In 1987, the existing detectors were only able to detect electron
anti-neutrinos but the now operating Super-Kamiokande and SNO experiments
will be capable of detecting Supernova $\nu _{e}$\ events. The analysis of
the energy distribution can in principle reveal or put limits on that kind
of effect.

The scenarios of neutrino mixing change considerably if one considers the
evidences from atmospheric and terrestrial neutrino experiments (for a
review see \cite{cald98}). The atmospheric neutrino anomaly and the zenith
angle dependence observed by Super-Kamiokande can be explained by $\nu _{\mu
}\rightarrow \nu _{\tau }$ oscillations\cite{suzu98}, $\nu _{\mu
}\rightarrow \nu _{e}$ excluded CHOOZ \cite{CHOOZ97}, but that is still
consistent with solar neutrinos oscillations into a linear combination of $%
\nu _{\mu }$ and $\nu _{\tau }$. That is no longer true if one takes in
consideration the LSND evidences of $\bar{\nu}_{\mu }\rightarrow \bar{\nu}%
_{e}$\ \cite{LSND96} and \ $\nu _{\mu }\rightarrow \nu _{e}$ \cite{LSND97}.
The much larger $\Delta m^{2}$ scale involved in LSND ($\Delta m_{e\mu
}^{2}>0.2\,{\rm eV}^{2}$), calls for a fourth flavor - a sterile neutrino -
into which solar $\nu _{e}$s should oscillate. That changes some features
concerning the effects of a NG field $\phi _{e}$ but one thing is
maintained: the signature of NG fields is a {\em surprise} {\em i.e.}, an
oscillation pattern not consistent with SM weak interactions \cite{bent98}.

An important point is that the Nambu-Goldstone fields are proportional to
the reaction rates that violate the relevant quantum numbers. In the case of
neutrino oscillations this manifests as a dependence on the neutrino
luminosity magnitudes. The effects of the Majoron fields if any, will be
observed in a shorter or longer interval of time depending on the actual
scale of lepton number symmetry breaking. Thus, the observation of such
correlation with the flux intensities, would thus provide a measurement of
the scale of spontaneous symmetry breaking.

\section{Resonant conversion of anti-neutrinos}

In the framework of the SM weak interactions the anti-neutrinos
cannot undergo a resonant oscillation if the neutrino masses follow the same
hierarchy as the charged leptons. That is because the SM potential $V_{W}$
is positive in an electron rich star medium and the potential reverses sign
for antineutrinos. The NG boson contributions are not only non-universal in
flavour but also their signs depend on the flavour content of the neutrino\
fluxes and dominant flavor transitions. This allows in principle other than
the standard MSW type of oscillations.

If one of the partial lepton numbers is spontaneously broken one expects the
same happens for all three flavours. Let $\sigma _{i}$ be the set of scalar
fields, singlets of SU(2)xU(1), with quantum numbers $\Lambda _{i}=({L_{e}},{%
L_{\mu }},{L_{\tau }})_{i}$ under the symmetry group ${\rm U(1)}_{e}\times 
{\rm U(1)}_{\mu }\times {\rm U(1)}_{\tau }$. If no residual symmetry is left
after spontaneous symmetry breaking there are three NG bosons. By going to
the physical basis as in Eqs.\ (\ref{tpsi}-\ref{tsigma}  ), now summing over the quantum
numbers $\Lambda ={L_{e}},{L_{\mu }},{L_{\tau }}$, 
\begin{eqnarray}
\chi ^{a} &=& e^{-i\,{\sum \,{\xi }}_{{{^{\Lambda }}}}{{\Lambda _{a}}}}\,\psi
^{a}\ ,\medskip \label{phys2a} \\
\sigma _{i} &=& e^{-i\,{\sum \,{\xi }}_{{{^{\Lambda }}}}{{\Lambda _{i}}}%
}\,\left( \left\langle \sigma _{i}\right\rangle +\rho _{i}\right) \ ,
\label{phys2b}
\end{eqnarray}
one obtains the Lagrangian of the NG\ fields 
\begin{equation}
{\cal L}={\frac{1}{2}}V_{\Lambda M}^{2}\,\partial _{\mu }\xi _{\Lambda
}\,\partial ^{\mu }\xi _{M}+{\cal L}_{{\rm int}}(\,\partial _{\mu }\xi
_{\Lambda })\,,  \label{lxi}
\end{equation}
the sum over $\Lambda ,M$ is implicit. The matrix 
\begin{equation}
V_{\Lambda M}^{2}=2\sum\limits_{i}{\Lambda _{i}}M_{i}\left| {\langle \sigma
_{i}\rangle }\right| ^{2}
\end{equation}
is in general non-diagonal, provided that some of the fields $\sigma _{i}$
have simultaneously more than one non-trivial quantum number. Such mixing
implies that the NG\ bosons mediate an action at a distance between two
distinct currents as show the equations of motion, 
\begin{equation}
\partial _{\mu }\partial ^{\mu }\xi _{\Lambda }=-\sum\limits_{M}G_{\Lambda
M}\,\partial _{\mu }J_{M}^{\mu }\,,  \label{ddxi}
\end{equation}
where $G_{\Lambda M}$ is the inverse matrix of $V_{\Lambda M}^{2}$.

In a stationary regime, if the neutrinos oscillate causing current
variations $\delta \vec{J}_{M}$, the expression for the gradients of the NG
bosons, $\vec{A}_{\Lambda }=-\vec{\nabla}\xi _{\Lambda }$, that generalizes
Eq.\ (\ref{Ae}) is \cite{bent97}: 
\begin{equation}
\vec{A}_{\Lambda }=-\sum\limits_{M}G_{\Lambda M}\,\delta \vec{J}_{M}\,.
\label{alv2}
\end{equation}
Since the interaction Lagrangian of the leptons has the same form as Eq.\ (%
\ref{lint}) {\em i.e.}, ${\cal L}_{{\rm int}}=$ $J_{\Lambda }^{\mu
}\,\partial _{\mu }\xi ^{\Lambda }$, the potential energy of a neutrino $\nu
_{\ell }$ subject to stationary NG fields is given by 
\begin{equation}
V_{\nu _{\ell }}=\vec{A}_{\ell }\cdot \vec{v}_{\nu }\;,  \label{Vnul}
\end{equation}
as in Eq.\ (\ref{Vnue}). Plugging in Eq.\ (\ref{alv2}), one gets for example, 
\[
V_{{\nu }_{\mu }}=-\left( G_{\mu e}\,\delta \vec{J}_{e}+G_{\mu \mu }\,\delta 
\vec{J}_{\mu }+G_{\mu \tau }\,\delta \vec{J}_{\tau }\right) \cdot \vec{v}%
_{\nu }\;. 
\]
This shows that the violation of a certain lepton number in one place
originates fields that acts on other flavors in other places.

In the old preferred scenario (before Super-Kamikande results on atmospheric
neutrinos \cite{suzu98}), the neutrino mixing angles are small, $\nu
_{e}\approx \nu _{1},$ $\nu _{\mu }\approx \nu _{2},$ $\nu _{\tau }\approx
\nu _{3}$, and the\ masses obey the hierarchy $m_{1}<m_{2}<m_{3}$. Then \cite
{beth86,walk87}, as the neutrinos come out of a Supernova the first resonant
transitions are between $\nu _{e}$ and $\nu _{\tau }$. It is outside their
resonance shell, at lower medium densities, that MSW transitions $\nu
_{e}\leftrightarrow {\nu _{\mu }}$, with smaller $\Delta m^{2}$, occur.

\strut In this meeting Atsuto Suzuki \cite{suzu98} reported Super-Kamiokande
results on the atmospheric neutrino anomaly and zenith angle dependence that
can be explained by $\nu _{\mu }\rightarrow \nu _{\tau }$ oscillations. The
best values, confirmed at Neutrino 98, are $\Delta m^{2}=5\times 10^{-3}\,%
{\rm eV}^{2}$ and $\sin ^{2}2\theta =1$. The alternative $\nu _{\mu
}\rightarrow \nu _{e}$ oscillation is excluded by CHOOZ \cite{CHOOZ97}. This
is consistent with a scenario where the heaviest of the neutrinos is
essentially a superposition, $\nu _{\rm{H}}$ (possibly maximal), of $\nu
_{\mu }$ and $\nu _{\tau }$; $\nu _{e}$ is the lightest neutrino and
oscillates in the Sun into the medium weight orthogonal superposition, $\nu
_{\rm{L}}$, of $\nu _{\mu }$ and $\nu _{\tau }$. In what follows what is
said about $\nu _{\tau }$, $\nu _{\mu }$ and quantum numbers ${L_{\tau }}$, $%
{L_{\mu }}$ can be maintained if replaced by $\nu _{\rm{H}}$, $\nu _{\rm{%
L}}$ and respective combinations of ${L_{\tau }}$ and ${L_{\mu }}$.

In a Supernova the $\nu _{e}$ flux is larger than the ${\nu _{\tau }}$ flux
and consequently more $\nu _{e}$ transform into ${\nu _{\tau }}$ than the
reverse which makes a net decrease of ${L_{e}}-{L_{\tau }}$. The current
densities $\vec{J}_{\tau }$ and $\vec{J}_{e}$ suffer the variations 
\begin{equation}
\delta \vec{J}_{\tau }=-\delta \vec{J}_{e}=\vec{j}(\nu _{e}\rightarrow {\nu
_{\tau }})-\vec{j}({\nu _{\tau }}\rightarrow \nu _{e})\,,  \label{djtau}
\end{equation}
that generate the NG fields $\xi _{e},\,\xi _{\mu },\,\xi _{\tau }$.
Applying Eq.\ (\ref{alv2}) one gets the vector potentials outside the sphere
where $\nu _{e}\leftrightarrow {\nu _{\tau }}$ takes place. The magnitude of
the constants $G_{\Lambda M}$ is directly related to the symmetry breaking
scales, but the flavor structure is arbitrary and is not constrained by the
neutrino\ mixing. The matrix $G_{\Lambda M}$ has only to be symmetric and
positive definite as demands the scalar bosons kinetic Lagrangian. Consider
the following hierarchy: $|G_{ee}|,|G_{e\mu }|,|G_{e\tau }|$ $<<|G_{\mu \tau
}|,G_{\tau \tau }$. Then, $\vec{A}_{e}$ is much smaller in magnitude than $%
\vec{A}_{\mu }$ and $\vec{A}_{\tau }$ and, assuming spherical symmetry, the
potential energies are well approximated by 
\begin{eqnarray}
V_{\nu _{e}} &=&V_{W}\ +A_{e}\ , \\
V_{{\nu }_{\mu }} &=&A_{\mu }\approx -G_{\mu \tau }\,\delta J_{\tau }\;,
\label{vnu2} \\
V_{{\nu _{\tau }}} &=&A_{\tau }\approx -G_{\tau \tau }\,\delta J_{\tau }\;, 
\end{eqnarray}
omitted the neutral current terms. The anti-neutrino\ potentials are the
symmetric of these.

The increment of $\tau $ lepton number per unity of time, $\delta F_{\tau }=%
\dot{N}({\nu _{e}\to \nu }_{{\tau }}{)-}\dot{N}({\nu _{\tau }\to \nu _{e})}$%
, is positive and so is the flux transfer $\delta\! J_{\tau }=\delta\! F_{\tau
}/4\pi r^{2}\!$. Defining $\delta\! F_{\tau }=\delta\! F_{51}\times
 \! $ $10^{51}%
{\rm erg/s/MeV}$ one estimates $\delta F_{51}=2-5\ $in the first 1/2
seconds of neutrino emission until the Supernova explosion and ${\rm \,}%
\delta F_{51}=0.2-0.5$\ at 1 second time. See \cite{bent97} for more
details. The NG potential for ${\nu _{\tau }}$ is negative but that is not
necessarily the case of ${\nu _{\mu }}$ because the element $G_{\mu \tau }$
can be negative! Using Eq.\ (\ref{VNG}),

\begin{equation}
A_{\mu }=-1.48\;\frac{G_{\mu \tau }^{F}}{G_{F}}\delta F_{51}\,\left( \frac{r%
}{10^{10}\,{\rm cm}^{{}}}\right) ^{-2}\times 10^{-12}\,\,{\rm eV}\;.
\end{equation}
The Schr\"{o}dinger equation for the anti-neutrinos is the same as for
neutrinos, Eq.\ (\ref{osc1}), but with a symmetric potential:

\begin{equation}
i\frac{\partial }{\partial \,r}\left( 
\begin{array}{c}
{\bar{\nu}_{e}} \\ 
\\ 
{\bar{\nu}_{\mu }}
\end{array}
\right) =\frac{1}{2E}\left( 
\begin{array}{cc}
-2E\,(V_{W}-A_{\mu }) & \quad {\rm \,}\frac{1}{2}\Delta m^{2}\sin 2\theta \\ 
&  \\ 
{\rm \,}\frac{1}{2}\Delta m^{2}\sin 2\theta & \quad {\rm \,\,}\Delta
m^{2}\cos 2\theta
\end{array}
\right) \left( 
\begin{array}{c}
{\bar{\nu}_{e}} \\ 
\\ 
{\bar{\nu}_{\mu }}
\end{array}
\right) \;,  \label{osc2}
\end{equation}
where $|A_{e}|\ll $ $|A_{\mu }|$ is implicit. Within the SM, the {effective
mass }$-2E\,V_{W}$ {would be negative,} and the resonance condition $%
-2E\,V_{W}=\Delta m^{2}\cos 2\theta $ could not be met. In presence of NG
fields, the effective mass {\ }$M_{e}^{2}=-2E\,(V_{W}-A_{\mu })$ is negative
at the high density regions, but since the NG potentials only decay with the
second power of the radius their contributions inevitably overcome at large
enough distances. That is a crucial point. If on top of that $G_{\mu \tau }$
is negative, {the effective mass }$M_{e}^{2}$ suddenly turns positive and
may reach the value $M_{\mu }^{2}=\Delta m^{2}\cos 2\theta $. In that case
there is a resonance layer where $M_{e}^{2}=M_{\mu }^{2}$ is met. In Fig.\ 4
it corresponds to the interception of the ''$e$'' and ''$\mu $'' curves, $%
M_{e}^{2}(r)$\ and $M_{\mu }^{2}$, respectively. The positive part of the
'potential' $M_{e}^{2}$ is plotted in Fig.\ 4 with values $G_{\mu \tau }=-G_{%
\rm{F}}$, ${\rm \,}\delta F_{51}=1$ or 3, $Y_{e}=1/2$, ${\rm \,}\tilde{M}%
_{31}=4$, and a neutrino\ energy of $20\,{\rm MeV}$, the same for all curves.%

\begin{figure}[t]
\centering
\includegraphics*[width=3.375in]{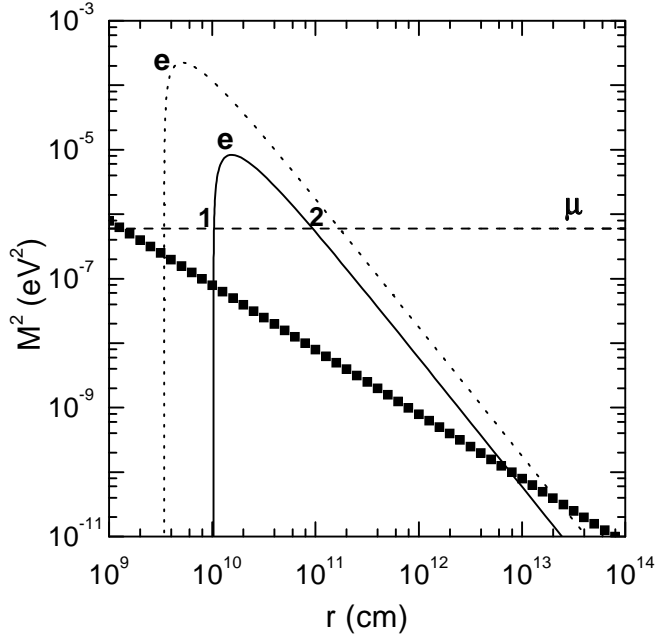}
\caption{$2E(V_{\bar\nu_e} -V_{\bar\nu_\mu} )$ is plotted in the solid and
dotted curves with $\delta F_{51} =1$ and 3 respectively.
The dashed line stands for a particular value of $\Delta m^{2}\cos 2\theta$
and the squares curve is the adiabatic boundary for 
$M_{\mu }^{2} \tan^2 2\theta$ in resonance 2.}
\end{figure}

Take ${\rm \,}M_{\mu }^{2}=6\times 10^{-7}\,{\rm eV}^{2}$ for example. The $%
M_{\mu }^{2}$ dashed line crosses the relative potential curve at two
points: the first, $r_{1}$, lies in the remarkably sharp transition from the 
$-V_{W}<0$ dominated region to the NG field dominated one and very likely a
non-adiabatic level crossing occurs there {\em i.e.}, ${\bar{\nu}_{e}}$
continues ${\bar{\nu}_{e}}$ and ${\bar{\nu}_{\mu }}$ continues ${\bar{\nu}%
_{\mu }}$. In the second point, $r_{2}$, the potential is smooth enough ($%
M_{e}^{2}\propto r^{-2}$) for adiabatic transitions. In general terms, the
conversion probability is enhanced \cite{mikh86,kuo89} if ${\rm \,}M_{\mu
}^{2}\,\tan ^{2}2\theta $ is larger than the rate $f(r)=|E\,{\rm d}V/V{\rm d}%
r|$ at resonance. In Fig.\ 4, $f(r)$ is plotted with squares for $V\propto
r^{-2}$. One can see that small mixing angle adiabatic conversions are
possible for a wide range of neutrino\ masses (for more details see \cite
{bent97}). Through this mechanism one could have resonant ${\bar{\nu}_{e}}%
\leftrightarrow {\bar{\nu}_{\mu }}$ oscillations in a Supernova whereas $\nu
_{e}\rightarrow {\nu _{\mu }}$ take place in the Sun. In particular, with
the parameters of the non-adiabatic solution \cite{haxt86,park86,rose86}
namely, $M_{\mu }^{2}\approx (3-12)\times 10^{-6}$$\,{\rm eV}^{2}$ and ${\rm %
\,}M_{\mu }^{2}\tan ^{2}2\theta \approx 4\times 10^{-8}\,{\rm eV}^{2}$,
provided that the neutrino\ fluxes are high enough (compare the dotted ${\rm %
\,}$and solid curves).

Since the predicted energy spectrum of ${\bar{\nu}_{\mu }}$ and ${\bar{\nu}%
_{\tau }}$ is harder than the ${\bar{\nu}_{e}}$ spectrum that kind of
oscillation could be manifest in the SN1987A events \cite{hira87}. Some
analyses made \cite{jege96,smir94}, but not all \cite{kern95},
disfavour the occurrence of ${\bar{\nu}_{e}}\leftrightarrow {\bar{\nu}_{\mu }%
}$ oscillations. This deserves some discussion \cite{bent97}, here I simply
stress that the effects caused by the NG fields strongly depend on magnitude
of the neutrino fluxes and on the scale of the global symmetry breaking as
well. Since the fluxes vary with time in a Supernova, one could have
anti-neutrino oscillations in the first instants of emission that fade away
when the fluxes drop below a certain threshold. The observation in the
future of such correlation between oscillation and flux magnitudes would
provide a measurement of the symmetry breaking scale.

\section{Conclusions}

We have shown that despite the fact that true Nambu-Goldstone (NG)\ bosons
only have derivative couplings, it is still possible to have coherent NG
fields. This happens whenever the quantum numbers they are associated with
are not conserved in reactions taking place in a large scale. If the lepton
numbers are spontaneously broken charges then, neutrinos oscillations in
stars produce classic NG fields. In addition, if the scale of symmetry
breaking is below 1 TeV, these fields are strong enough in Supernovae to
change the neutrino flavor dynamics.

The observable signatures are:

\noindent - Time variation of neutrino oscillation patterns in correlation
with the flux magnitudes. \noindent This is a consequence of the NG fields
being proportional to the time rate of reactions and therefore to the
neutrino fluxes.

\noindent - Incongruence of Supernova neutrino oscillation patterns with the
evidence obtained from solar, atmosferic and terrestrial neutrinos. The NG
fields are not expected to be significant in these cases even if they are in
Supernovae because the fluxes are much smaller. This is ultimately a result
of the derivative coupling nature of the NG bosons.

Finally, if we will are lucky enough to see such kind of phenomena in the
future, then, Supernova neutrinos will provide a measurement of the scale of
lepton number symmetry breaking.

\strut 

\section*{Acknowledgements}

This work was supported in part by the project ESO/P/PRO/1127/96.

\end{document}